# Role of magnetism and electron-lattice interactions in STM spectra of cuprates

J. Hwang[1], T. Timusk[1,2], J.P. Carbotte[1,2]
[1] Department of Physics and Astronomy, McMaster University, Hamilton, ON L8S 4M1, Canada, [2] The Canadian Institute of Advanced Research, Toronto, Ontario M5G 1Z8, Canada

The study of bosonic modes that couple to the charge carriers is a key element in understanding superconductivity. Lee et al. have used atomic-resolution scanning tunneling spectroscopy (STM) in an attempt to extract the spectrum of of these modes[1] in the high temperature superconductor $Bi_2Sr_2CaCu_3O_8$ and they found a mode whose frequency did not depend on doping but changed on $^{16}O$ to $^{18}O$ isotopic substitution leading them to suggest a role for lattice modes (phonons). A careful examination of their published data reveals a weaker but distinct feature that has all the characteristics of the magnetic excitation identified as the bosonic mode in other competing experiments[2-4]. We suggest the lattice mode seen by Lee et al. is not relevant to superconductivity and is due to inelasting tunneling through the insulating oxide layer[5].

Tunneling spectroscopy was used successfully in the conventional low temperature (BCS) superconductors to extract a spectrum which agreed in detail with the phonons found in the same materials by neutron spectroscopy. The method has also been applied to high temperature superconductors, by Zasadzinsky[2,3] using SIS break junction tunneling spectroscopy. These authors found a sharp bosonic peak whose spectroscopic properties agreed with those of the well known neutron resonance: the frequency of the mode was proportional to the superconducting transition temperature and the mode strength decreased markedly with doping. Similar doping dependence of the mode was observed by Hwang et al. using optical spectroscopy[4]. We should note here that while scanning tunneling superconductor-insulator-normal metal (SIN) spectroscopy gives a direct measure of the density of states of the superconducting state, superconductor-insulator-superconductor (SIS) break junction tunneling spectroscopy produces a convolution of the two superconducting densities of states which however can be analyzed to yield a bosonic spectrum.

A plausible explanation of the observations of Lee et al. was offered in a recent paper by Pilgram et al.[5] where they suggested that the phonon Lee et al. found was the result of the direct excitation of the apical oxygen vibrations and not density of states effects associated with the superconducting $CuO_2$ layer. Here we propose a mechanism that reconciles the two different sets of tunneling experiments as well as optics with the suggestion of Pilgram et al.. We note that superconductivity gives a negative peak in the second derivative of the tunneling current[6] and a positive one for the inelastic tunneling mechanism[7]. The fact that Lee et al. observed a positive peak does not rule out a contribution from the superconducting plane since in SIN tunneling there is an additional positive structure on the high energy side of the main negative peak[6].

A careful examination of Fig. 4b of Lee et al. shows that the histograms of the maxima of $d^2I/dV^2$ are asymmetric and consist of two components: a sharp, narrow doping independent peak and a second, broad component whose position and strength vary with doping. The procedure used by the authors to plot these curves would be expected to lose all spectroscopic information about the bosonic spectral function except for the frequency of the maximum point. However the doping dependent asymmetry in Fig. 4b shows that this is not the case. There are several possible mechanisms that might reveal weaker spectral features in the histograms. For example, in the presence of sufficient noise, through dithering, weaker features will be selected from time to time. Or surface inhomogeneities might favor one process over the other in different parts of the sample. As shown in Fig 1. data of Lee et al. are well fit by a sum of two gaussians, one centered at 52 meV the other with a variable center frequency. The amplitude of the broad peak decreases with doping and the center frequency follows $T_c$, being lowest for the highly underdoped and highly overdoped samples. Panel f) shows a plot of the position of the broad component, normalized to $\Delta$ vs. $\Delta$ along with tunneling data from

Zasadzinski et al.[2] with which it agrees. These results are, as pointed out in ref. 2, consistent with a magnetic resonance model for the peak.

Finally it is interesting to note that for the most overdoped sample, shown as a black curve, the broad component is very weak. This is consistent with the proposal made by Hwang et al. (from optics) that the magnetic resonance contribution to the self energy weakens in the overdoped region for $T_c$ < 50 K in qualitative agreement with SIS tunneling. Also, as a final note, the reason that the spectra of Zasadzinski et al. and Hwang et al. do not show the phonon contribution is because they probe the properties of the conducting $CuO_2$ plane while, in the light of the proposal of Pilgram et al., STM spectra contain a contribution from inelastic tunneling through the SrO layer. Our analysis shows that the STM technique can also see the bosonic spectrum that controls the self energy of the superconducting carriers and, as pointed out by Scalapino in the related News and Views article, may lead us to the identification of magnetic fluctuations as the 'pairing glue' responsible for high temperature superconductivity[8].

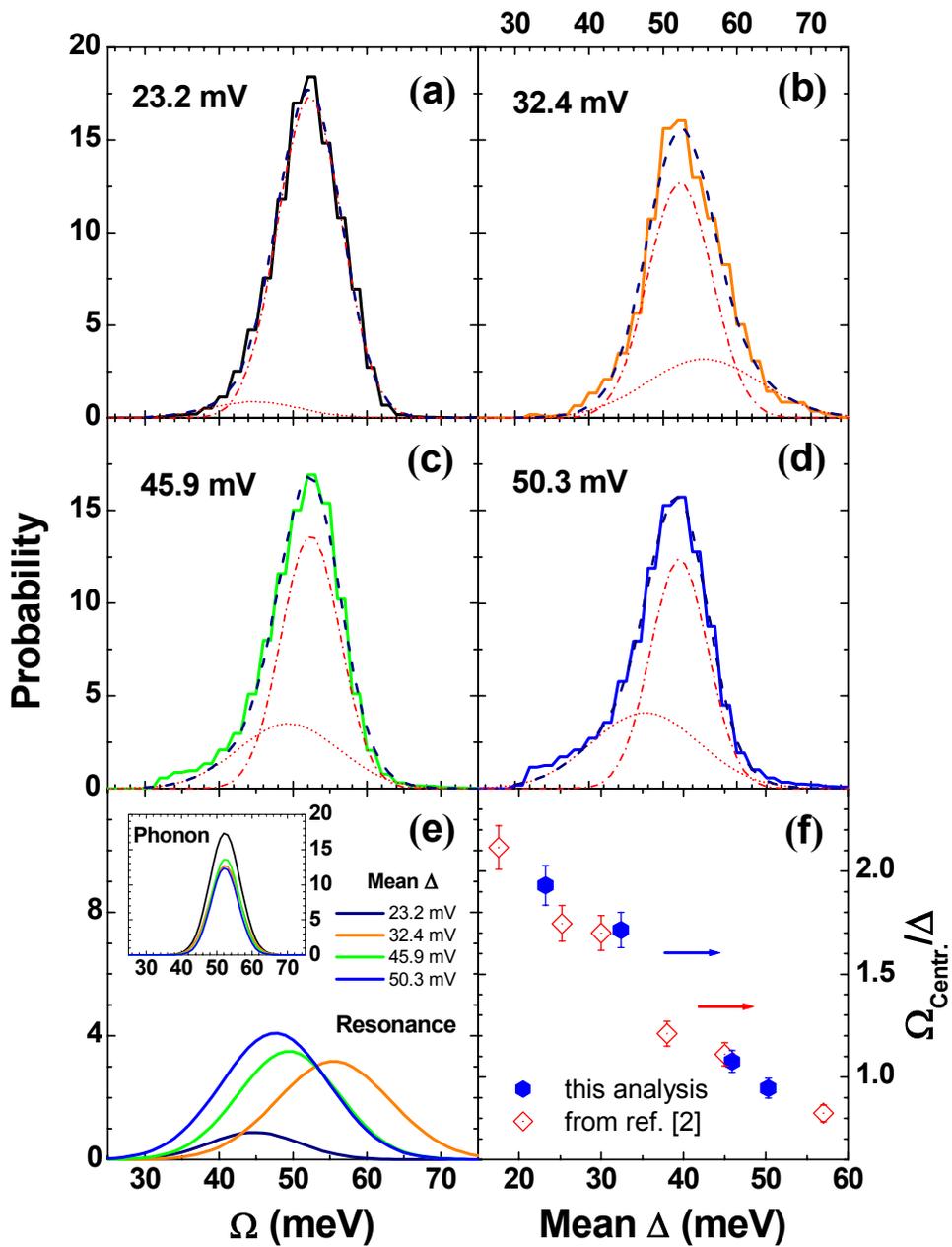

<LEGEND>Panels a) to d), the bosonic peak reproduced from Lee *et al.* for samples of different doping levels from the most overdoped a) to the most underdoped d) labeled by the corresponding value of the superconducting gap. The dashed curves are a least squares fit to a two-component model, a sharp peak fixed at a frequency of 52 meV and a broader component allowed to vary in frequency and width. Panel e) shows the two components of the fit and panel f) compares the center frequency of the broad component, normalized to $\Delta$, with data from break junction tunneling, ref. (2). For a better comparison we used the bias voltage, which gives the maximum positive slope in the peak-dip-hump structure of the tunneling conductance, as $\Omega_{Centr.}$ instead of the bias voltage at the dip.